\documentclass[runningheads]{llncs}

\usepackage[a4paper, total={6.5in, 10in}]{geometry}
\usepackage[style=apa , backend=biber]{biblatex}
\usepackage{setspace}
\usepackage{graphicx}
\usepackage{array}

\begin{document}

%%%%%%%%%%%%%%%%
% TITLE & NAME %
%%%%%%%%%%%%%%%%
% \selectlanguage{portuguese}  % <-- ACTION: UNCOMMENT for english or portuguese
%\selectlanguage{english}  % <-- ACTION: UNCOMMENT for english or portuguese
\title{Large Language Models to Enhance Business Process Modeling: Past, Present, and Future Trends}
%\title{Leveraging Large Language Models to Enhance Business Process Modeling: A Literature Review} % <-- ACTION: WRITE your project title here

\titlerunning{LLMs to enhance Business Process Modeling}
\author{João Bettencourt\inst{1} \and
Sérgio Guerreiro\inst{1,2}\orcidID{0000-0002-8627-3338}}
\authorrunning{Bettencourt et al.}
\institute{INESC-ID, R. Alves Redol 9, 1000-029 Lisbon, Portugal
\and
Instituto Superior Técnico, Universidade de Lisboa, Av. Rovisco Pais 1, 1049-001 Lisboa, Portugal\\
\email{\{joao.pais.bettencourt,sergio.guerreiro\}@tecnico.ulisboa.pt}}

%<-- ACTION: WRITE your complete name here
%\istid{96880} % <-- ACTION: WRITE your student number  here (your student number, not the IST-ID)
%\email{joao.pais.bettencourt@tecnico.ulisboa.pt} % <-- ACTION: WRITE your e-mail here
%\advisors{Sérgio Guerreiro} % <-- ACTION: WRITE your advisor name here

% To make the title
\maketitle
%\thispagestyle{empty}

%%%%%%%%%%%%
% ABSTRACT %
%%%%%%%%%%%%
% To remove indentation before abstract
%\setlength{\abstitleskip}{-\absparindent}
\begin{abstract}

\textbf{Purpose:}
Recent advances in Generative Artificial Intelligence, particularly Large Language Models (LLMs), have stimulated growing interest in automating or assisting Business Process Modeling tasks using natural language. Several approaches have been proposed to transform textual process descriptions into BPMN and related workflow models. However, the extent to which these approaches effectively support complex process modeling in organizational settings remains unclear. 
This article presents a literature review of AI-driven methods for transforming natural language into BPMN process models, with a particular focus on the role of LLMs.

\textbf{Design/methodology/approach:}
Following a structured review strategy, relevant studies were identified and analyzed to classify existing approaches, examine how LLMs are integrated into text-to-model pipelines, and investigate the evaluation practices used to assess generated models.

\textbf{Findings:}
The analysis reveals a clear shift from rule-based and traditional NLP pipelines toward LLM-based architectures that rely on prompt engineering, intermediate representations, and iterative refinement mechanisms. 
While these approaches significantly expand the capabilities of automated process model generation, the literature also exposes persistent challenges related to semantic correctness, evaluation fragmentation, reproducibility, and limited validation in real-world organizational contexts.
 
\textbf{Originality:}
Based on these findings, this review identifies key research gaps and discusses promising directions for future research, including the integration of contextual knowledge through Retrieval-Augmented Generation (RAG), its integration with LLMs, the development of interactive modeling architectures, and the need for more comprehensive and standardized evaluation frameworks.

% \keywords{Process Modeling, Business Process Management, Generative AI, Large Language Models, Retrieval-Augmented Generation} % <--- ACTION: WRITE YOUR KEYWORDS HERE

% \textbf{Keywords ---} Process Modeling, Business Process Management, Generative AI, Large Language Models, Retrieval-Augmented Generation


%\clearpage
\end{abstract}

%%%%%%%%%%%%%%%%%%%%%
% TABLE OF CONTENTS %
%%%%%%%%%%%%%%%%%%%%%
%\tableofcontents
%\thispagestyle{empty}
%\clearpage

%% <-- ACTION: ADD/EDIT/REMOVE/COMMENT SECTIONS IF YOU NEED TO
%%%%%%%%%%%%%%%%
% INTRODUCTION %
%%%%%%%%%%%%%%%%

\section{Introduction}

% Parágrafo 1 — O problema
Business Process Modeling is a core activity within the field of Business Process Management (BPM), enabling organizations to document, analyze, implement, and improve their operational processes. As described by~\textcite{Dumas2018}, process modeling plays a central role in BPM initiatives and is commonly realized using standardized notations such as Business Process Model and Notation (BPMN)~\parencite{bpmn2014}. Despite its widespread adoption, process modeling is cognitively demanding and a time-consuming task, typically requiring specialized expertise and extensive interaction with stakeholders and documentation. As a result, process models are often outdated, incomplete, or misaligned with actual operational practices, which limits their practical value in real-world settings.

% Parágrafo 2 — Importância + RQs implícitas
Recent advances in Generative Artificial Intelligence (AI), specifically Large Language Models (LLMs), have sparked growing interest in automating or assisting process modeling tasks from natural language specification. A growing body of work has explored different AI with LLM-based approaches for transforming textual process descriptions into BPMN models. This evolution raises fundamental questions about whether such models can effectively support business process modeling in practice, especially in organizational environments characterized by fragmented documentation, implicit domain knowledge, and evolving requirements. In this context, ~\textcite{Bennoit2024} emphasize the need to understand how LLMs can be meaningfully integrated into BPM activities across the BPM lifecycle, highlighting the usability and the contextual awareness as key challenges demanding further research.

% Parágrafo 3 — Porque é difícil (por que abordagens ingênuas falham)
However, applying LLMs to business process modeling is inherently challenging. Naive approaches — such as rule-based NLP techniques or prompt-driven LLM usage without contextual enrichment — tend to frame process modeling as a predominantly transformation-based task grounded solely in the initial textual input. While such approaches may produce structurally coherent models, they remain constrained by the information explicitly provided in the description and fail to address the inherent ambiguity of natural language, the need for semantic interpretation and reasoning, and the structural rigor required by process modeling notations~\parencite{guerreiro2021conceptualizing}. These limitations are exacerbated in organizational settings, where complex process knowledge is distributed across heterogeneous documents, informal practices, and evolving contexts that cannot be inferred from isolated textual inputs alone. This distinction between syntactic validity and semantic correctness is well recognized in the literature, where automatically generated BPMN models may conform to the formal rules of the notation, but fails to accurately capture the intended process meaning~\parencite{Hoerner2026}.

% Parágrafo 4 — Porque não foi resolvido antes / limitações do estado da arte
Existing research has investigated the use of LLMs for transforming natural language into process models, as well as interactive assistants that support human modelers, as shown in recent studies by~\cite{Kourani2025}, and~\cite{Ziche2024}. While these works report encouraging results, they address different research objectives and exhibit specific limitations.~\cite{Kourani2025} primarily evaluate LLM capabilities within a framework whose internal representation constrains the class of process models that can be expressed; but when applied to more complex modeling contexts, performance decreases, which is consistent with their goal of benchmarking LLM capabilities rather than supporting modeling as an iterative activity. In the opposite perspective,~\textcite{Ziche2024} propose an interactive assistant, but their evaluation relies largely on subjective user assessment, limiting the empirical rigour of their conclusions. Moreover, several transformation-oriented approaches adopt a predominantly one-shot generation paradigm, often emphasizing the production of sound models by construction. While this design choice ensures structural correctness, it provides more limited support for iterative refinement scenarios in which models are progressively evolving from multiple natural language inputs. In such settings, explicit mechanisms for preserving context and user intent across iterations are typically absent, as exemplified by frameworks such as GIVUP, developed in~\textcite{Nivon2025}. These gaps are partly explained by the relative novelty of LLM-based architectures in the BPM domain, where mechanisms for incorporating external knowledge and contextual information are still underexplored. In this regard,~\textcite{Azeredo2025} analyzes Retrieval-Augmented Generation (RAG) as a means to improve semantic grounding and contextual consistency, but without addressing its integration into Business Process Modeling workflows.

% Parágrafo 5 — Problema resumido, proposta, resultados esperados e limitações
The central problem examined in this paper is the extent to which LLMs can effectively support Business Process Modeling in organizations, where process knowledge is fragmented, implicit, and distributed across heterogeneous sources. While recent advances demonstrate the feasibility of generating BPMN models from natural language, the related literature reveals structural limitations that extend beyond isolated technical shortcomings. This paper classify and synthesizes existing approaches, critically analyzes their methodological and practical constraints, and identifies persistent gaps related to semantic consistency, contextual grounding, iterative refinement, and evaluation rigor. Building on this analysis, the discussion formulates research venues that position Retrieval-Augmented Generation (RAG), contextual continuity mechanisms, and structured evaluation protocols as promising directions for advancing the field.

This paper is organized as follows.
Section~\ref{sec.literaturereview} presents the results of a literature review done in the context of LLMs and BPM.
Afterwards Section~\ref{sec:discussion} discusses the results obtained in the reviewed literature.
Then, Section~\ref{sec.openresearch} identifies the research venues in this field.
Finally, Section~\ref{sec.conclusions} concludes the paper.

%%%%%%%%%%%%%%%%
% Related work %
%%%%%%%%%%%%%%%%

\section{Literature Review}
\label{sec.literaturereview}

This paper employs a Literature Review (LR) guided by the methodological principles proposed by~\textcite{Kitchenham2007}. Although these guidelines are originally defined for conducting Systematic Literature Reviews (SLRs), the present study does not claim to constitute a fully systematic review due to the limited number of relevant primary studies that are available at present time. Nevertheless, the structured procedures recommended by Kitchenham were followed to ensure rigor, transparency, and reproducibility in the search, selection, and analysis of the literature.

The objective of this review is to identify, analyse, and synthesise existing scientific knowledge on AI-driven transformation from natural language into BPMN, and other forms (if any), of related process models.

Accordingly to Kitchenham, structured LRs provide a rigorous and reproducible means of identifying and evaluating relevant research. The review process is organised into three main phases:

\begin{enumerate}
    \item \textbf{Planning}: This phase begins by establishing the motivation for conducting the LR, after which the Research Questions (RQs) are defined to guide the remainder of the review.
    
    \item \textbf{Conducting}: This phase comprises a structured search for primary studies across selected scientific databases, followed by study selection based on the predefined criteria. The selected studies form the basis of this LR.
    
    \item \textbf{Reporting}: In the final phase, the extracted data is synthesised and analysed to answer the RQs and the results are reported in a structured manner.
\end{enumerate}

The outcomes of this LR provide a comprehensive overview of existing approaches, the role of LLMs, evaluation practices, comparative insights, and the challenges and open gaps identified in the literature. 

\subsection{Planning Phase}

\subsubsection{Motivation}

BPMN plays a central role in the analysis, implementation, communication, automation and improvement of organizational processes~\parencite{milani2025business}. It is focused on the activities, resources, flows, gateways, events, messages, and (basic) data objects that occur in a business process. Its widespread adoption stems from several factors: it provides a standardized and expressive way to represent control-flow logic and data~\parencite{lopes2023assessing}; it enables shared understanding among technical and non-technical stakeholders; and it serves as a foundation for downstream tasks such as simulation, compliance checking, and process automation. Despite its importance, deriving BPMN models from textual descriptions remains a difficult and labor-intensive activity. Manual modeling is often time-consuming, inconsistent, and prone to omissions or ambiguities that exist in natural language descriptions. Conversely, Automated approaches have historically struggled with the limitations of traditional Natural Language Processing (NLP) methods, which frequently fail to capture implicit logic, solve complex dependencies, or generalize beyond narrow rule-based pipelines~\parencite{beerepoot2023biggest}. These issues contribute to deliver low-quality models and hinder large-scale or real-time applications.

The recent capabilities of LLMs raise the possibility of addressing many of these longstanding challenges. Their ability to understand context, reason over sequences of events, and generate structured outputs suggests that they could support, or even automate, the transformation of natural language descriptions into BPMN or related workflow representations. However, it remains unclear how LLMs are currently being used for this task, how they compare to previous AI-based or rule-based approaches, and what evaluation methods are appropriate for assessing their performance.

For these reasons, a LR was conducted to map the existing landscape. The review aims to identify (i) approaches that use AI or NLP to transform natural language into BPMN or other workflow diagrams, (ii) how LLMs are being integrated into this transformation process, (iii) the evaluation methods used to assess generated models, (iv) how LLM-based techniques compare to earlier methods, and (v) the challenges, limitations, and research gaps that remain. Together, these insights provide the foundation for understanding the current state of the field and for informing subsequent design and methodological decisions in this work.

\subsubsection{Research Questions}

This review adopts an exploratory stance to develop a broad and well-supported understanding of the current landscape of AI-driven transformations from natural language into BPMN and related workflow models. Both conceptual advances and practical implementations are considered, as understanding the field requires examining not only what has been proposed, but also how these approaches operate and are evaluated in practice.

According to the guidelines and the work of~\cite{Kitchenham2007}, a LR should clarify its methodological orientation. Two closely related forms exist: the Systematic Literature Review, which aims to synthesize evidence to answer specific research questions, and the Systematic Mapping Study, which provides a broader overview of how research is distributed across a domain. Since AI-assisted text-to-BPMN generation is still an emerging area, this work adopts a hybrid strategy, combining elements of both approaches.

To guide the investigation, five RQs were formulated based on the objectives established in the Motivation section:

\begin{itemize}
    \item \textbf{RQ1 – Approaches:} What methods using AI have been proposed to transform natural language into BPMN models?
    \item \textbf{RQ2 – Role of LLMs:} How are LLMs being used in this transformation process?
    \item \textbf{RQ3 – Evaluation:} How are the proposed approaches evaluated?
    \item \textbf{RQ4 – Comparison:} How do LLM-based approaches compare with earlier NLP-based or rule-based methods for text-to-BPMN generation?
    \item \textbf{RQ5 – Challenges and Gaps:} What challenges, limitations, and open research gaps are reported in the literature?
\end{itemize}

In the context of this hybrid design, RQ1, RQ2, and RQ5 align more closely with the goals of a Systematic Mapping Study, as they aim to identify, describe, and categorise existing research. Conversely, RQ3 and RQ4 follow the perspective of a Systematic Literature Review, focusing on aggregating evidence regarding evaluation practices and comparing current approaches with prior methods.

\subsection{Conducting}

\subsubsection{Search Process}

The search for relevant literature began with the definition of a comprehensive search string. To ensure the retrieval of studies addressing the RQs, keywords were grouped into four main categories: translation, AI methods, business process domain, and BPMN notation. Each group was designed to capture different aspects of the topic:

\begin{enumerate}
    \item \textbf{Translation (process of conversion):} 
    \begin{itemize}
        \item translation
        \item "text-to-model"
        \item "natural language"
        \item "textual description"
    \end{itemize}
    \item \textbf{AI (methods and technology):}
    \begin{itemize}
        \item AI
        \item "artificial intelligence"
        \item "machine learning"
        \item "deep learning"
        \item "large language model" OR LLM OR GPT OR ChatGPT OR transformer
        \item "natural language processing" OR NLP
    \end{itemize}
    \item \textbf{Business Process (domain):}
    \begin{itemize}
        \item "business process"
        \item "business process modeling"
        \item "workflow"
        \item "process diagram"
    \end{itemize}
    \item \textbf{BPMN (notation):}
    \begin{itemize}
        \item BPMN
        \item "Business Process Model and Notation"
    \end{itemize}
\end{enumerate}

This resulted in the \textbf{Final Search String}:

\begin{verbatim}
("translation" OR "natural language" OR "text-to-model" OR "textual description") 
AND ("AI" OR "artificial intelligence" OR "machine learning" OR "deep learning" 
     OR "large language model" OR LLM OR GPT OR ChatGPT OR "transformer" 
     OR "natural language processing" OR NLP) 
AND ("business process" OR "business process modeling" OR "workflow" OR "process diagram") 
AND (BPMN OR "Business Process Model and Notation")
\end{verbatim}

The search was conducted across three major digital libraries to ensure comprehensive coverage of the literature: Web of Science (WoS), IEEE Xplore and ACM Digital Library. A publication date filter was applied to include only studies from 2015 onwards.

\subsubsection{Exclusion Criteria}

The following exclusion criteria (EC) were applied to filter the retrieved literature:

\begin{itemize}
    \item \textbf{EC1 – Access:} The full text of the paper is not accessible.
    \item \textbf{EC2 – Language:} The paper is not written in English.
    \item \textbf{EC3 – Domain Relevance:} The study is not related to BPM, BPMN, BPMN generation, or similar process modeling tasks.
    \item \textbf{EC4 – Input/Method/Output Relevance:} The study neither applies AI, NLP, or LLM techniques for text-to-model transformation, nor produces or analyses BPMN diagrams or other structured process models as output. Papers focusing on intermediate steps, such as entity or element extraction from text, are considered relevant and are not excluded under this criterion.
    \item \textbf{EC5 – Research Questions Relevance:} The study does not provide information relevant to any of the research questions (RQ1–RQ5).
\end{itemize}

These criteria, particularly EC4, follow an Input-Method-Output perspective:  
\begin{itemize}
    \item \textbf{Input:} Natural language or textual process descriptions.  
    \item \textbf{Method:} Any type of AI, including machine learning, deep learning, or LLM-based approaches.  
    \item \textbf{Output:} Structured process models, such as BPMN diagrams, workflow models, or other control-flow-oriented representations.
\end{itemize}

Although the search string included the term BPMN, which is the primary target output of the approaches considered in this LR, methods generating other types of \textbf{control-flow–oriented process models} were also included. Alternative modeling languages, such as Entity–Relationship models, UML class diagrams for data modeling, or UML statecharts for state-based modeling, are not included in this review, as they serve different purposes. This broader scope aligns with the approach taken by~\textcite{SchulerAlpers2024}, and helps capture methods that could potentially be adapted or extended for BPMN generation.

From initial readings of the literature, it became evident that many methods involve an initial transformation or extraction phase, where information is converted into an intermediate structured representation before being transformed into the final BPMN model. As a result, the exclusion criteria were adjusted to ensure that papers addressing only part of this pipeline—such as studies focusing on the intermediate extraction or structuring phase—were still considered relevant and included in the review, even if they do not cover the entire transformation from natural language to BPMN.

\subsubsection{Initial Analysis / Abstract Reading}
\label{sec:abstract-reading}

In the first screening phase, the title and abstract of each retrieved paper were examined to determine whether it should proceed to the next stage of the review. For most papers, these two elements were sufficient to make an initial decision. 

Papers that clearly met one or more exclusion criteria were immediately eliminated, while those whose relevance could not be fully determined from the title and abstract were retained for further assessment in the next phase. For each excluded paper, the corresponding EC was recorded to ensure transparency in the selection process.

\subsubsection{In-Depth Analysis / Full Reading}
\label{sec:full-reading}

Following the initial screening, the selected papers underwent a full, in-depth analysis. This phase served both to eliminate studies for which the initial abstract review was inconclusive and to gain a comprehensive understanding of the contributions of each document. 
During this process, each paper was read thoroughly, and notes were taken on key findings relevant to the research questions. 
%After completing this detailed review, a total of 41 studies were retained for the final analysis.

\subsubsection{Results}

\begin{figure}
  \centering
  \includegraphics[width=\textwidth]{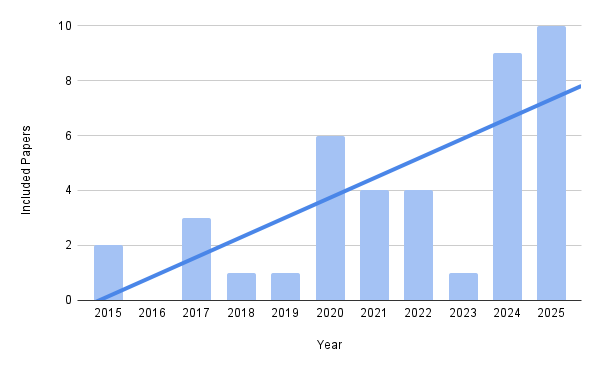}
  \caption{The included number of paper organized by year (from authors).}
  \label{fig:Included results per year}
\end{figure}

Applying the search string described earlier across the three digital libraries yielded 211 unique results. Following Sections~\ref{sec:abstract-reading} and~\ref{sec:full-reading}, and considering the defined exclusion criteria, a total of 41 papers were retained for the review.  
Among these, 3 were SLR or LR, 4 were conceptual studies (\emph{i.e.}, they proposed approaches but had not yet implemented or evaluated them), and the remaining 34 were experimental studies.
The distribution of included papers per year is shown in Figure~\ref{fig:Included results per year}. An obvious increase is noted in the recent years.
All extracted information, including synthesis across digital libraries, deduplication, and categorization, is publicly available in the Final Data Extraction and Manipulation Sheet at~\parencite{Bettencourt2026}.

\subsection{Reporting}

\subsubsection{RQ1 – Approaches: What methods using AI have been proposed to transform natural language into BPMN models?}
\label{sec:rq1_approaches}

AI is a broad umbrella term encompassing a wide range of techniques and paradigms. Although all included studies employ AI-based methods, it is necessary to clarify how AI is framed in this review in order to support a consistent and transparent categorization of the proposed approaches.

Given the focus of this research on LLMs and related techniques, the identified methods were classified into two main categories: Generative AI (GenAI) and non-Generative AI (NoGenAI). This classification follows the conceptualization and taxonomy proposed by~\textcite{Hafner2025}, reflecting how GenAI technologies are operationalized and discussed within the BPM literature.

Accordingly, GenAI is understood as encompassing AI technologies that enable the automated creation, transformation, or optimization of process-related artefacts, as well as advanced language-based process understanding. As a result, all AI approaches considered as GenAI by~\textcite{Hafner2025} are classified as GenAI in this review. This includes transformer-based models such as BERT, which are therefore treated as GenAI despite not being generative in a strict machine learning sense. This choice ensures conceptual alignment with prior BPM research and supports a consistent categorization of the reviewed studies.

Many of the reviewed papers employ multi-step pipelines to transform natural language input into BPMN models. For classification purposes, a study was assigned to the GenAI category if at least one step in its pipeline made use of a GenAI technology as defined above, even if other steps relied on NoGenAI techniques. Only the 34 experimental studies were included in this categorization.

\subsubsection*{Non-Generative AI Approaches}

Starting with the approaches that do not use Generative AI, these methods primarily rely on traditional NLP techniques. While some NoGenAI solutions use Machine Learning (ML) or, in a few cases, Deep Learning (DL), the most common approaches fall under Symbolic AI, including rule-based and pattern-matching methods. These approaches typically follow a multi-step pipeline composed of the following phases:

\begin{itemize}
    \item \textbf{Text Preprocessing and Syntactic Analysis:} 
    Parsing, tokenization with PoS tagging, dependency parsing, and extraction of Subject-Verb-Object (SVO) constructs to identify grammatical structures and process elements.

    \item \textbf{Semantic Analysis and Disambiguation:} 
    Extracting meaning and resolving ambiguity using lexical resources, semantic role labeling (SRL), and coreference resolution.

    \item \textbf{Rule Application and Intermediate Representation:}  
    In Symbolic AI approaches, this phase consists of applying explicit, predefined mapping rules to linguistic patterns, producing a formal intermediate representation (such as CREWS rules, spreadsheet-based structures, or fact types) that can later be transformed into BPMN elements.  
    In ML and feature-based DL approaches, explicit rule application is replaced by supervised predictive models, which learn to map syntactic and semantic features generated in earlier phases to process elements and relations. These models output structured representations — such as labeled entities or inferred relations — that serve the same role as the symbolic intermediate model.
\end{itemize}

\subsubsection*{Generative AI Approaches}

Within the GenAI category, the same taxonomy introduced by~\textcite{Hafner2025} is further applied to structure the identified approaches. According to this taxonomy, the most relevant GenAI technologies in the BPM context are Generative Adversarial Networks (GANs), Variational Autoencoders (VAEs), and Transformer-based Models.

GANs are primarily applied to tasks such as image generation, anomaly detection, and data augmentation. VAEs, on the other hand, are commonly used for anomaly detection, generative modeling, and recommendation systems. Since these approaches are generally not linked to NLP, they appear rarely in the reviewed studies; in fact, only one instance of a VAE was identified, reported by~\textcite{Qiu2025}. In that study,~\textcite{Qiu2025} incorporated VAEs within an Automated Process Generation module to generate diverse variants of business processes. The workflow functioned as follows:

\begin{itemize}
    \item \textbf{Learning process structure:} Long Short-Term Memory networks (LSTMs) and Graph Neural Networks (GNNs) were first used to learn the structure of existing processes from historical data.
    \item \textbf{Generating variants:} The VAEs then generated multiple plausible process variants based on the learned structures.
    \item \textbf{Optimization:} These variants were subsequently evaluated and optimized—using Reinforcement Learning or other criteria—to identify the most efficient or high-quality process designs.
\end{itemize}

In short, VAEs enabled the system to explore a wide range of potential process configurations, while the optimization methods selected the best-performing ones.

Transformer-based models clearly dominate the GenAI approaches, consistent with the findings of~\textcite{Hafner2025}, where LLMs constitute the vast majority of generative AI technologies applied in process modeling and extraction. The specific roles and applications of LLMs in transforming natural language into workflow models, such as BPMN, will be examined in the following subsection.

\begin{figure}
  \centering
  \includegraphics[width=\textwidth]{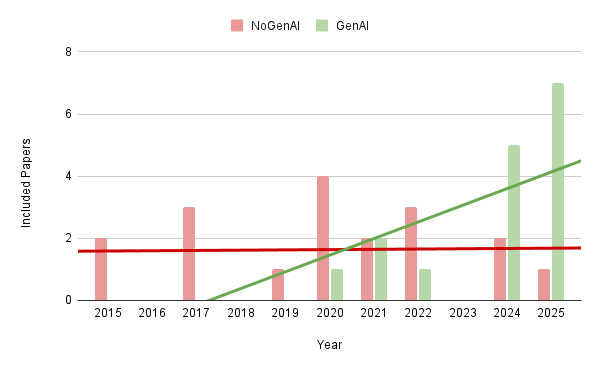}  \caption{Included results per year categorized. An increase in generative AI is noticed, while noGenAI is not evolving (from authors).}
  \label{fig:included-results-per-year-gen-ai-linear}
  %% change this to GenAI and NoGenAI
\end{figure}

Figure~\ref{fig:included-results-per-year-gen-ai-linear} illustrates the yearly distribution of accepted papers, separated into NoGenAI and GenAI approaches. While early years are dominated by NoGenAI methods, GenAI solutions first appear in 2020 and by 2024–2025, GenAI clearly surpasses NoGenAI, indicating the growing adoption of generative approaches.

\subsubsection{RQ2 – Role of LLMs: How are large language models (LLMs) being used in this transformation process?}

Unlike earlier NLP or rule-based methods, which operate through predefined linguistic rules or supervised feature mappings, LLMs interpret text using internal representations learned from large-scale corpora. These representations capture statistical and semantic patterns of language, allowing LLMs to understand relationships between words, events, and actions without relying on handcrafted rules. As a result, they can extract, synthesize, and formalize process information directly from natural language descriptions. Their role within text-to-BPMN pipelines is multifaceted, often spanning information extraction, structured generation, intermediate modeling, and iterative refinement.

LLMs extract process-relevant information from unstructured text, identifying activities, actors, control-flow dependencies, decision points, and process constraints. Rather than directly generating BPMN diagrams, they produce structured intermediate representations (formal models, DSLs, JSON schemas, or code templates) that can be validated, programmatically interpreted, or deterministically converted into BPMN. In the following sections, the instruction of LLMs for process modeling is explored, addressing the key question of how task-specific guidance is provided.

\subsubsection*{Prompt Engineering}

The effectiveness of LLM-based approaches in text-to-BPMN transformation depends not only on the underlying model, but crucially on how the model is instructed. Since LLMs are general-purpose systems, most studies adapt them to the process modeling task through \textbf{Prompt Engineering}, which explicitly structures the model’s behavior using carefully designed instructions. Prompt engineering provides the mechanism through which task-specific knowledge, modeling constraints, and output expectations are communicated to the LLM without modifying its parameters. This enables reproducible customization across different LLM architectures.

\textcite{Neuberger2025} propose a prompting structure specifically designed for \textit{extracting process model information from natural language descriptions}, which combines three core components:

\begin{itemize}
    \item \textbf{Context:} Specifies the overall objective of the task, such as extracting workflow information or adhering to BPMN modeling conventions.
    \item \textbf{Task Description:} Defines the types of elements to be extracted, the semantic or syntactic constraints to be followed, and the \textbf{meta-language} used to represent the extracted process information.
    \item \textbf{Restrictions:} Enforces formal output requirements, such as valid JSON schemas, well-formed domain-specific languages, or structurally sound intermediate models (e.g., POWL).
\end{itemize}

\subsubsection*{In-Context Learning}

Several studies rely on \textbf{In-Context Learning (ICL)}, where the LLM is instructed through examples embedded directly in the prompt. In this setting, \textbf{few-shot prompting} provides pairs of natural language process descriptions and their corresponding structured representations, demonstrating how process-relevant information should be extracted and encoded~\parencite{Neuberger2025, Kourani2025, Ziche2024, Bellan2022, Kogler2024}. This approach mitigates the limited task-specific knowledge of LLMs without requiring parameter updates.~\textcite{Kourani2025} further employ \textbf{negative prompting} to explicitly discourage invalid constructs and improve adherence to modeling constraints.

\subsubsection*{Knowledge Injection and RAG}

Beyond example-based instruction, some approaches provide guidance through \textbf{Knowledge Injection}, supplying conceptual or structural information rather than demonstrations. This includes the specification of intermediate language semantics by~\textcite{Kourani2025}, formal definitions of process elements and relations proposed by~\textcite{Bellan2022}, task-specific meta-languages that define which elements to extract introduced by~\textcite{Neuberger2025}, DSL grammars enriched with natural language descriptions to support correct text-to-model mapping presented by~\textcite{Kogler2024}, and the injection of ontological modeling principles—such as universal and existential quantifiers—through instruction panels available during verification tasks described by~\textcite{Tsaneva2024}.

In addition, Retrieval-Augmented Generation (RAG) enables the use of external information retrieved at inference time and incorporated into the prompt. These sources may include documentation, glossaries, regulations, or structured knowledge such as graphs or ontologies representing domain concepts and their relationships.~\textcite{Ziche2024} employ RAG to incorporate internal organizational documentation and process modeling rules, enabling the LLM to adapt its responses to the specific operational context. Beyond the BPM domain, retrieval-augmented approaches have also been studied as a general mechanism for improving semantic grounding and contextual consistency in document-centric systems, as analyzed by~\textcite{Azeredo2025}. By grounding generation in up-to-date and context-specific information, RAG enables more informed and adaptive process modeling and has been discussed by~\textcite{Bennoit2024}, as a mechanism for extending LLMs toward continuously informed BPM support tools.

\subsubsection*{Domain Adaptation and Fine-Tuning}

While most approaches rely exclusively on prompting, a smaller subset of studies explores fine-tuning to specialize LLMs for process modeling tasks. In these cases, models are adapted using curated datasets that pair process descriptions with ground-truth representations, resulting in more stable, domain-aware, and structurally compliant outputs.

Fine-tuning is typically adopted under the following conditions:

\begin{itemize}
    \item \textbf{Specialized domains:} where process descriptions rely on domain-specific terminology or conventions that are insufficiently covered by general-purpose models. This includes sector-specific adaptations such as Lumina BPM proposed by~\textcite{Li2025}, fine-tuned for electric power workflows, ProcessLLM introduced by~\textcite{Buss2025}, trained on BPM manuals and modeling guidelines, TeProM developed by~\textcite{Sun2025}, specialized on SAP operational documents, and ERNIE-based approaches for administrative and regulatory processes presented by~\textcite{Qiu2025}.
    
    \item \textbf{Strict intermediate representations:} when the target formalism requires precise syntactic and structural compliance. Examples include fine-tuning for structured DSLs such as PiperFlow in Lumina BPM proposed by~\textcite{Li2025}, JSON- or CSV-based extraction pipelines for BPMN rendering introduced by~\textcite{Ajmal2024}, XML-compliant BPMN generation in ProcessLLM implemented by~\textcite{Buss2025}, and logic-based representations such as LTL formulas in GIVUP presented by~\textcite{Nivon2025}.
    
    \item \textbf{Limitations of prompting:} in settings where prompting alone yields inconsistent structures, shallow semantic interpretations, or recurrent hallucinations. Several studies explicitly motivate fine-tuning to address these limitations, including ProcessLLM developed by~\textcite{Buss2025}, GIVUP proposed by~\textcite{Nivon2025}, TeProM introduced by~\textcite{Sun2025}, and automated BPMN generation approaches reported by~\textcite{Ajmal2024} which struggle with anaphora resolution and grammatical consistency under prompt-only setups.
\end{itemize}

While more resource-intensive than prompting-based approaches, fine-tuned LLMs consistently demonstrate higher reliability and stronger adherence to modeling conventions in complex or domain-specific process modeling scenarios.

\subsubsection*{Intermediate Representations and Code Generation}

LLMs typically do not produce BPMN diagrams directly. Instead, they generate textual or structural intermediates that are easier to validate and transform. These intermediates typically take one of the following forms:

\begin{itemize}
    \item \textbf{Formal modeling languages:} such as POWL, which ensure soundness and compositionality.~\textcite{Kourani2025} guide the LLM to generate POWL models, which are then deterministically converted into executable BPMN.

    \item \textbf{Domain-specific DSLs:} like PiperFlow or the Sketch Miner textual syntax.~\textcite{Li2025} describe how the Lumina BPM framework outputs PiperFlow sequences from procedural text, which are later transformed into BPMN diagrams. Similarly, ~\textcite{Ivanchikj2020} use the Sketch Miner syntax to allow users to instantly visualize BPMN models from simplified textual descriptions.

    \item \textbf{Structured schemas:} usually JSON-based representations of activities, relations, or gateways.~\textcite{Kogler2024} translate natural language process descriptions into a JSON-based DSL, which is validated against a strict schema and then converted into executable Java code.

    \item \textbf{Code templates:} where the LLM generates Python or Java instructions that construct BPMN models using predefined libraries.~\textcite{Kourani2025} describe how the model outputs scripts calling functions such as \texttt{gen.\allowbreak activity()} or \texttt{gen.\allowbreak xor()} to build the process programmatically, with the Python script generating the corresponding POWL models.
\end{itemize}

These intermediate representations act as the operational interface between the LLM and the downstream modeling engine. They allow deterministic systems to enforce correctness, detect syntactic errors, and translate the LLM’s generative output into executable or visual BPMN models.

\subsubsection*{Iterative Refinement and Error Handling}

In process modeling pipelines, LLMs are commonly integrated into \textit{iterative refinement loops}. LLMs do not inherently refine outputs; this behavior relies on feedback-driven system architectures. In such settings, downstream components validate the LLM’s output and return diagnostic feedback — for example, malformed JSON, invalid POWL constructs, or missing dependencies — which is then used to guide subsequent generations. 

Refinement mechanisms in these architectures typically include:

\begin{itemize}
    \item \textbf{Automated feedback loops:} where deterministic validators detect syntactic or structural errors and return explicit diagnostics to the LLM for correction.~\textcite{Kourani2025} describe how the ProMoAI framework implements an automated error-handling loop that captures Python execution failures or POWL validation errors and feeds explicit error messages back to the LLM, allowing up to 15 correction iterations.

    \item \textbf{Self-correction strategies:} such as self-critique, in which the model revises its own output based on task constraints or evaluation criteria.~\textcite{Kourani2025} instruct the LLM to critically reassess its initial model against the original process description to identify logical improvements, while ~\textcite{Neuberger2025} propose reflection-based prompting that requires the model to justify its outputs, helping to surface ambiguities and hallucinations.

    \item \textbf{Candidate generation and selection:} where multiple alternative outputs are produced and ranked, either by the model itself or by external scoring functions.~\textcite{Kourani2025} describe how the system generates several alternative process models from the same description and uses LLM-based self-evaluation to select the variant that best satisfies conformity criteria.

    \item \textbf{Human-in-the-loop refinement:} allowing users to iteratively query, correct, or clarify specific parts of the process model through conversational interaction.~\textcite{Ziche2024} present an operational setting where human modelers continuously review and refine chatbot outputs,~\textcite{Kogler2024} enable users to visually edit models and provide additional natural language instructions for further refinement, and~\textcite{Bellan2022} explore incremental extraction through a sequence of guided question–answer interactions.
\end{itemize}

Together, these mechanisms form a feedback-driven architectural design that supports not only initial generation, but also the iterative refinement, debugging, and improvement of intermediate process representations before their final transformation into executable or visual BPMN artifacts.

\subsubsection*{Main Takeaways}

The approaches discussed above are not mutually exclusive. In practice, many studies combine multiple techniques—such as prompt engineering, in-context learning, knowledge injection, and fine-tuning—to guide LLMs effectively and achieve the final process modeling results. This combination leverages the strengths of each method to produce accurate, structured, and semantically sound BPMN representations.

\subsubsection{RQ3 – Evaluation: How are LLM-based approaches evaluated?}
\label{sec:rq3-evaluation}

Evaluation practices in LLM-based business process modeling vary significantly across the literature. This diversity reflects differences in the scope and objectives of the proposed approaches, as well as in the types of outputs they aim to produce. Rather than relying on a single evaluation methodology, existing work adopts strategies that align with the specific goals of each system.

Broadly speaking, three evaluation patterns can be identified.

\paragraph{1) Benchmark-driven evaluation under controlled modeling assumptions.}
Some frameworks aim to benchmark and compare the capabilities of LLMs for process modeling under controlled conditions. A representative example is ProMoAI, initially introduced by~\textcite{Kourani2024} and further developed in subsequent work~\parencite{Kourani2025}. ProMoAI evaluates LLMs using a curated benchmark consisting of textual process descriptions paired with ground-truth models expressed in an intermediate language (POWL). The evaluation relies on behavior-oriented conformance checking, where event logs generated from the reference models are replayed on the generated models to compute behavioral measures such as fitness and precision. Under this controlled setup, the reported results indicate high levels of behavioral validity.

\paragraph{2) Evaluation of practical generation tools on heterogeneous descriptions.}
Other approaches focus on generating BPMN models from more heterogeneous descriptions, closer to those encountered in practice. In such contexts, evaluation often relies less on strict automated metrics and more on expert-based assessments of correctness. This scope is illustrated by the approach proposed by~\textcite{Nivon2025}, whose evaluation considers textual descriptions originating from multiple sources and evaluates generated models using categories such as \emph{valid}, \emph{ambiguous}, or \emph{incorrect}. Additional indicators such as execution time are also reported.

\paragraph{3) System-level evaluation for human-centered modeling support.}
A third group of approaches evaluates LLM-based systems as tools intended to assist human modelers. In these cases, evaluation focuses on usability, feasibility, and modeling support rather than purely automated correctness metrics. This perspective is exemplified by the PRODIGY system proposed by~\textcite{Ziche2024}, whose evaluation is conducted through a qualitative case study involving enterprise users interacting with the system.

\paragraph{Summary.}
Overall, evaluation in LLM-based business process modeling reflects the objectives of each approach. Benchmark-oriented frameworks emphasize controlled experimentation and automated behavioral metrics, practical generation tools rely on expert-based validation of generated models, and human-centered systems are evaluated through qualitative case studies.

\subsubsection{RQ4 – Comparison: How do LLM-based approaches compare with earlier NLP- or rule-based methods for BPMN generation?}

The reviewed literature highlights a fundamental transition from rigid, manually engineered pipelines toward flexible, data-driven, and conversational architectures. Rather than incrementally improving earlier NLP techniques, LLM-based approaches redefine how natural language is interpreted and transformed into BPMN models.

\subsubsection*{Methodological Shift}

Earlier NLP- and rule-based approaches are typically organized as pipes-and-filters pipelines, combining syntactic parsing, semantic role labeling, and manually defined extraction rules. These systems assume relatively stable grammatical patterns and are often tuned to narrow domains, which limits their robustness in open-ended or ambiguous texts.

In contrast, LLM-based approaches rely primarily on \emph{prompt engineering} and \emph{in-context learning}. This enables rapid adaptation of general-purpose models to BPM tasks without the need for extensive rule engineering. Instead of producing intermediate linguistic annotations, LLMs often generate structured representations (e.g., JSON or XML) directly from raw text, which are then translated into BPMN by downstream components.

\subsubsection*{Capabilities and Scalability}

A key advantage of LLM-based approaches lies in their ability to process and aggregate information across large volumes of heterogeneous data, going beyond the single-document and pattern-constrained scope of earlier NLP pipelines. In particular, LLM-based systems demonstrate:

\begin{itemize}
    \item \textbf{Multi-source integration:} Ability to combine process information distributed across heterogeneous artifacts such as e-mails, manuals, and chat logs.
    \item \textbf{Process variant detection:} Capability to identify alternative execution paths, exceptions, and process variants that emerge across different descriptions or sources.
    \item \textbf{Iterative model refinement:} Support for incremental, exploratory refinement of process models through repeated natural-language interactions and feedback loops.
\end{itemize}

In addition, the \textbf{conversational interaction paradigm} of LLMs significantly lowers the technical entry barrier, allowing non-expert users to create, query, and refine process models using natural language rather than formal modeling notations or specialized BPM tools.

\subsubsection*{Precision, Reliability, and Control}

Rule-based systems are deterministic and transparent: given the same input, they always produce the same output, and errors can typically be traced back to specific rules. This predictability, however, comes at the cost of limited semantic coverage and poor generalization.

LLMs exhibit stronger contextual and semantic understanding, but this flexibility introduces uncertainty. Hallucinated activities, incorrect relations, or inconsistent abstractions are frequently reported, making \emph{human-in-the-loop validation} a recurring requirement. As a result, LLM-based approaches trade strict predictability for expressive power.

\subsubsection*{Engineering Effort and Preparation}

Traditional approaches require significant upfront investment in \textbf{rule design}, \textbf{tuning}, and \textbf{long-term maintenance}, typically involving domain experts and business analysts. While LLM-based approaches substantially reduce the need for manual rule coding, they do not eliminate engineering effort entirely. Instead, the effort shifts toward \textbf{prompt and context design}, the definition of \textbf{iterative refinement strategies}, and the implementation of \textbf{validation and correction mechanisms} to ensure structural and semantic reliability.

\begin{table}[h]
\centering
\caption{Comparison between earlier NLP/rule-based approaches and LLM-based approaches.}
\setlength\extrarowheight{6pt}
\begin{tabular}{p{4cm} p{5cm} p{5cm}}
\hline
\textbf{Dimension} & \textbf{Earlier NLP / Rule-Based} & \textbf{LLM-Based} \\
\hline
Modeling & Rule-driven pipelines & Prompt-guided, data-driven \\
Input & Structured or simple text & Open-ended, heterogeneous text \\
Interaction & Analyst-driven, offline & Conversational, iterative \\
Reliability & Deterministic but rigid & Expressive but may hallucinate \\
Engineering & Manual rule design & Prompt design and validation \\
Scalability & Domain-limited & Cross-domain, large-scale \\
\hline
\end{tabular}
\end{table}

\subsubsection{RQ5 – Challenges and Gaps: What challenges, limitations, and open research gaps are reported in the literature?}

The automated transformation of natural language into BPMN models using LLMs faces multiple challenges, spanning technical limitations, data constraints, and methodological gaps.

\subsubsection*{Technical Limitations}
\begin{itemize}
    \item \textbf{LLM-specific issues:} hallucinations, limited reasoning capabilities, sensitivity to prompt design, and difficulty enforcing structured output.
    \item \textbf{Linguistic challenges:} ambiguity, active vs. passive voice, pronoun resolution, and long-distance dependencies that complicate process extraction.
    \item \textbf{Residual NLP challenges:} while LLMs outperform classical NLP pipelines, complex control flows (XOR, AND, loops) and fine-grained semantic nuances can still be problematic.
\end{itemize}

\subsubsection*{Data and Input Challenges}

LLM-based BPMN generation is strongly affected by the quality and availability of input data. \textbf{Data scarcity} is a key issue because training and evaluating LLMs require large, high-quality, human-annotated datasets, which are often unavailable. \textbf{Heterogeneity and ambiguity} in process descriptions—arising from differences in writing style, structure, and terminology—make it difficult for models to extract consistent process elements across sources. Finally, \textbf{privacy and access constraints} limit the use of external LLM services due to confidential business data, while the token costs of commercial APIs can hinder large-scale experimentation and adoption.

\subsubsection*{Methodological and Research Gaps}
\begin{itemize}
    \item \textbf{Fragmented evaluation:} absence of standardized metrics, benchmarks, and systematic comparisons.
    \item \textbf{Human-in-the-loop integration:} need for iterative refinement and expert validation to manage hallucinations and ensure correctness.
    \item \textbf{Unexplored modalities:} limited investigation of images, voice, or multimedia as input sources for LLM-based BPM generation.
    \item \textbf{Real-world validation:} most studies remain theoretical or small-scale; comprehensive evaluation in enterprise environments is still lacking.
\end{itemize}

\subsubsection*{Main Takeaway}
LLM-based approaches are moving the field toward flexible, data-driven, and interactive BPMN generation, but their reliability, dependency on quality data, and fragmented evaluation practices highlight the need for robust, standardized, and human-guided frameworks.

%%%%%%%%%%%%%%%%%
% Discussion %
%%%%%%%%%%%%%%%%%
\section{Results Discussion}
\label{sec:discussion}

The analysis conducted in this paper reveals a rapidly evolving landscape of AI-driven approaches for transforming natural language into BPMN and related process models. While LLMs have substantially expanded the expressive and semantic capabilities of text-to-model pipelines, the literature consistently exposes limitations that go beyond isolated technical shortcomings.

Across the reviewed studies, certain structural and semantic limitations are repeatedly acknowledged. However, when these findings are examined collectively, additional methodological and evaluative patterns emerge that warrant closer scrutiny. The following discussion synthesizes both the commonly reported constraints and the broader structural implications identified through this review.

\subsection{Structural and Semantic Limitations of LLM-Generated BPMNs}

A recurring theme across the reviewed studies concerns the persistence of hallucinations and semantic inconsistencies, particularly when process descriptions involve complex control-flow structures, implicit dependencies, or ambiguous terminology. Although some approaches succeed in producing syntactically valid BPMN models—often through intermediate representations such as POWL—semantic correctness remains fragile.

Many approaches implicitly adopt an input-centric modeling paradigm, in which the textual description is treated as the primary and often exclusive source of process knowledge. This framing assumes that the provided text is sufficiently self-contained to support accurate model construction. However, real organizational settings rarely conform to this premise. Process knowledge is typically fragmented across heterogeneous documents, embedded in informal practices, and shaped by implicit domain assumptions. Terminology may vary between departments, and modeling decisions frequently require contextual interpretation rather than literal translation.

Additionally, several frameworks adopt a predominantly one-shot generation paradigm. While suitable for benchmarking and controlled experimentation, this approach contrasts with the inherently iterative nature of human process modeling. Modeling is rarely a single-pass activity; it evolves through clarification, negotiation, and refinement. When iterative mechanisms are implemented, they are often system-driven correction loops rather than modeling-driven refinements guided by evolving user intent. This raises a subtle trade-off: delegating contextual memory management to the system reduces user effort but may also reduce transparency and control over how the evolving model state is constructed.

As a result, generated models may achieve structural coherence and formal soundness while remaining semantically misaligned with the intended organizational process. Moreover, when modeling is reduced to a predominantly one-shot generation task, the broader modeling activity itself is abstracted away from its inherently iterative, interpretative, and collaborative nature. In such settings, the system optimizes for output production rather than for supporting the progressive construction of shared process understanding. This exposes a central tension in the current state of the art: \textbf{automatic generation does not necessarily equate to organizationally meaningful modeling, nor does it adequately replicate the modeling process through which such meaning is constructed}.

\subsection{Methodological Fragilities}

While structural and semantic limitations are widely acknowledged in the literature, a closer examination of the reviewed studies reveals additional methodological fragilities that are less systematically articulated. When considered collectively, these patterns raise concerns regarding the robustness, comparability, and interpretability of reported findings.

\subsubsection{Reproducibility}

Reproducibility emerges as a particularly salient concern. Many approaches rely on publicly accessible LLM APIs whose internal configurations evolve over time. Model updates, undocumented parameter adjustments, and version changes introduce volatility into experimental setups. Consequently, results that appear robust at a given point in time may become difficult—or even impossible—to replicate later under nominally similar conditions.

One potential mitigation strategy involves the use of locally deployable Small(er) Language Models (SLMs), which offer greater experimental control and stability. However, the adequacy of such models remains debated.~\textcite{Lauer2026} report that several small LLMs with fewer than seven billion parameters consistently failed to generate valid BPMN-XML models in their experiments. At the same time, the authors acknowledge that alternative process model representations—such as JSON-based formats—may be more accessible for language. This observation suggests that the limitations observed for smaller models may be closely tied to the requirement of directly generating BPMN-XML, which imposes strict structural and syntactic constraints. When intermediate representations are employed instead, the structural burden on the model may be reduced. This leaves open the empirical question of whether SLMs could provide sufficient quality under architectures that decouple semantic extraction from final BPMN serialization.

Reproducibility concerns are further intertwined with enterprise-level privacy and security constraints. Organizations may be unwilling—or legally unable—to expose sensitive documentation to external LLM services, limiting the feasibility of API-dependent architectures and complicating experimental transparency. This tension is illustrated in the case study by~\textcite{Ziche2024}, where the proposed system relies on a non–privacy-compliant external LLM service and explicitly acknowledges constraints regarding organizational applicability. Such dependencies affect not only deployment feasibility but also the generalizability of evaluation results.

\subsubsection{Training Data Realism}

A second fragility concerns the realism and representativeness of training data. Fine-tuned approaches typically rely on curated datasets composed of textual process descriptions paired with corresponding ground-truth models. While such datasets enable supervised learning and controlled experimentation, they are often limited in size and structural complexity, frequently involving relatively simple control-flow patterns, restricted gateway combinations, and shallow nesting structures. Consequently, models are predominantly trained on structurally simple and unambiguous examples, raising concerns about their ability to generalize effectively to the ambiguity and complexity of real organizational processes.

\subsection{Evaluation}

While Section~\ref{sec:rq3-evaluation} describes the different evaluation strategies adopted in the literature, a closer examination of these practices reveals several limitations that complicate the interpretation and comparability of reported results.

As in the case of training data, one recurring issue concerns the construction of benchmark datasets used for automated evaluation. In some cases, textual descriptions and their corresponding process models are intentionally co-designed to ensure compatibility with the modeling formalism employed by the framework. A representative example is the ProMoAI benchmark proposed by~\textcite{Kourani2025}, where both the textual descriptions and the reference models are constrained to structures expressible in the intermediate language POWL. This design enables systematic behavioral evaluation through conformance checking and allows metrics such as fitness and precision to be computed automatically.

However, this setup also restricts the modeling space. By avoiding structures that fall outside the expressiveness of the intermediate representation, the benchmark reduces both linguistic and structural ambiguity in the input descriptions. As a result, strong performance under these conditions should be interpreted in light of the constraints imposed by the benchmark itself.

This limitation becomes more apparent when the same approach is used as a comparison baseline under different evaluation conditions. For example, when ProMoAI is used within the broader experimental setting of the GIVUP framework proposed by~\textcite{Nivon2025}, the reported validity rates drop to around 50\%, substantially lower than the values reported in its original benchmark. In contrast, GIVUP reports approximately 83\% valid models on its own datasets. This discrepancy reflects differences in evaluation scope rather than contradictory findings. In ProMoAI, both the textual descriptions and the reference models are designed to remain within the expressiveness of the intermediate representation (POWL), enabling systematic benchmarking of LLM capabilities under controlled modeling assumptions. When evaluated in more heterogeneous contexts with fewer structural constraints, such as those considered in GIVUP, the same generation strategies face a substantially more complex modeling environment. This does not undermine the validity of the ProMoAI benchmark; rather, it highlights that its primary objective is to compare the capabilities of LLMs under controlled conditions, rather than to develop a framework intended to support process modeling in open and heterogeneous settings.

At the same time, the evaluation design of GIVUP introduces other challenges. Because the framework performs one-shot model generation and evaluates outputs using expert-defined categories such as \emph{valid}, \emph{ambiguous}, or \emph{incorrect}, the results depend heavily on human interpretation. This approach captures aspects of practical modeling that automated benchmarks may miss, but it also makes direct comparisons with other approaches difficult, particularly with systems that incorporate iterative refinement or human-in-the-loop modeling workflows.

A different set of limitations appears in system-level evaluations such as the PRODIGY framework proposed by~\textcite{Ziche2024}. In this case, the evaluation focuses on the perceived usefulness of the system when assisting enterprise process modelers. The reported results are based on qualitative observations from a case study involving organizational users rather than on quantitative comparisons against ground-truth models. While such evaluations provide valuable insights into real-world applicability, they remain inherently subjective and do not isolate the contribution of specific architectural components. Notably, although PRODIGY employs RAG to incorporate organizational knowledge, the evaluation does not measure the impact of RAG independently from the overall system behavior. As a result, it remains unclear to what extent the observed benefits arise from RAG itself or from the broader interaction design of the system.

Taken together, these observations highlight a structural challenge for evaluation in LLM-based business process modeling. Benchmark-oriented studies emphasize automated behavioral metrics under constrained modeling assumptions, while practical tools and interactive systems rely on expert judgment or qualitative case studies. Although each strategy aligns with the goals of the respective approaches, the resulting evaluation landscape remains fragmented. Consequently, results reported across different frameworks are often difficult to compare directly, and strong performance within a given evaluation setup does not necessarily translate to broader modeling contexts.

Recent work has attempted to address some of these evaluation challenges through more comprehensive benchmarking frameworks. A notable example is the BEF4LLM framework proposed by~\textcite{Lauer2026}, which introduces a structured evaluation methodology combining multiple perspectives of process model quality. In contrast to many earlier evaluations that focus primarily on control-flow correctness, BEF4LLM attempts to capture broader aspects of model quality, including syntactic validity, behavioral correctness, and pragmatic aspects of the resulting models. The framework also mitigates some common evaluation pitfalls—for instance, by incorporating semantic similarity when comparing activity labels, allowing synonymous expressions to be recognized as valid matches rather than requiring exact textual correspondence.

Nevertheless, the framework itself acknowledges several limitations. The experiments rely on a relatively small dataset of only 105 text–BPMN pairs, which constrains the generalizability of the reported results. Furthermore, some aspects remain outside the current scope of the framework, such as additional evaluation metrics related to iterative generation processes (\emph{e.g.}, cyclicity metrics) and LLM-specific performance indicators like inference time or memory consumption. Certain evaluation dimensions may also penalize complex process models, as larger BPMN models tend to receive lower pragmatic scores than simpler ones. Despite these limitations, BEF4LLM currently represents one of the most comprehensive attempts to establish a systematic and multi-perspective evaluation methodology for LLM-based process model generation.

%\subsection{Research Hypotheses}

%The analysis suggests that the core research challenge lies not in marginally improving generation accuracy, but in integrating contextual awareness, iterative refinement, methodological rigor, and reproducibility into LLM-based BPM systems.

%From this perspective, the following investigation hypotheses emerge:

%\begin{itemize}
%    \item \textbf{H1:} The integration of Retrieval-Augmented Generation (RAG) improves the semantic consistency of BPMNs generated from natural language.
%    \item \textbf{H2:} Interactive systems with preserved contextual memory reduce inconsistencies across iterative refinements.
%    \item \textbf{H3:} Locally controlled or experimentally stable language models enhance reproducibility in BPMN generation tasks.
%    \item \textbf{H4:} Structured and standardized evaluation protocols provide more reliable assessments of semantic model quality than isolated case studies.
%\end{itemize}

%A particularly relevant open question concerns the nature of the documents supplied to RAG mechanisms. Should retrieval focus on modeling guidelines to improve structural compliance, on internal organizational documentation to enhance contextual grounding, or on a hybrid combination of both? Existing implementations incorporating RAG remain exploratory, and its isolated impact on semantic consistency has not yet been rigorously quantified.

%%%%%%%%%%%%%%%%%
% ResearchVenues %
%%%%%%%%%%%%%%%%%
\section{Research Challenges}
\label{sec.openresearch}

The research in automated generation of complete business process models, and the possibility of covering all types of (required) interactions to decrease the possibility of deviations is not new, \emph{e.g.}, the discussion introduced about this issue back in~\textcite{winograd1987understanding}.
However, today, the challenge is how to design solutions that can cope with the growing complexity of business processes. Many companies are struggling with digitization transformation requiring the design of hundreds of business processes and expect to consume as little time and resources as possible. For instance, it is a challenge to deal with the cascading effect in interactions among different revocations in process executions. 

On the one hand, the problem's complexity scales significantly when business processes are analyzed not merely through their control flow, but through their intricate interactions with data~\parencite{Dumas2018,gianola2023verification}. Indeed, a huge body of research has been recently devoted to the problem of integrating data and processes to achieve a more comprehensive understanding of their concrete interplay within BPM~\parencite{calvanese2019formal}. This requires investigation of how data influence the process behavior, and how the control flow of the process impacts on the data it queries and manipulates. 
It is also remarkable the result of~\textcite{genon2010analysing} reporting that even with classical training it is difficult that (Human) business analysts produce models that are coherent and semantically rich, and conversely, that the same models cannot be easily understood by (Human) non-technical or (Human) non-analyst stakeholders.

On the other hand, it is well-known that BPMN is the \emph{de facto} standard used by industry and researchers for business process modeling and execution. However, BPMN has several pitfalls, it requires the usage of a complex language that uses too many different symbols (over 200 graphic symbols), the lack of a formal scientific and theoretical ground on the meaning of its symbols, and a method for its use~\parencite{dijkman2008semantics}. The results is that BPMN can lead to disparate model interpretations. ~\textcite{fickinger2013construct} concluded that BPMN has a level of 51,3\% of overlapping language concepts and lacks state concept to ensure a sounder semantics. Even the efforts to standardize the BPMN usage are short in guaranteeing the models quality~\parencite{laue2010visualization}. The ambiguity is even higher when data dimensions are taken into consideration, such as the one of case variables carrying data objects. BPMN is completely agnostic regarding the type of data supported, since it does not provide any constraints/restrictions or formal interpretation/semantics on the nature of the usable data types~\parencite{gianola2023verification}.

Taking in consideration these two perspectives, it is identifiable the need for new solutions that can co-assist the generation of complex processes, including cascading processes, and not-only happy flows patterns. In turn, the lack of the BPMN formalism is a problem for creating datasets to be used directly by a generative AI solutions.

In detail, BPMN can be distinguished in intended for human interpretation (usual presented as visual representation) and the intended for machine interpretation (in XML format). It is of key importance to define which datasets will be used for AI training and for RAG. Without this definition, the results obtained will not be aligned with the expected expressiveness. 

%This paper identifies the opportunity of using enriched BPMN machine interpretation for LLM fine-tuning and RAG graph database definition, and also including the visualization part of the BPMN definition. More specifically, the training datasets need to be curated to include the expected visualization concerns.

Since 2024, Transformer-based models clearly dominate the Generative AI approaches in process modeling and extraction. LLMs extract process-relevant information from unstructured text, identify activities, actors, control-flow dependencies, decision points, and process constraints. Rather than directly generating BPMN diagrams, they can produce structured intermediate representations (formal models, DSLs, JSON schemas, or code templates) that can be validated, programmatically interpreted, or deterministically converted into BPMN. This has already been discussed in~\textcite{dumas2023ai} and many contributions are competing in this research thread. Yet, many hallucinations and semantic inconsistencies are still identified.

RAG plays an important role of organizing the knowledge offered by the business processes ontology into a graph database that could be consumed by LLM. This representation allows the contextualization of patterns for business processes, leveraging a solution that is not only based on the transformer engine – mainly based on language – but enhanced with tested business processes good practices that are compatible with human communication theory, \emph{e.g.},~\parencite{dietz2024enterprise}. 

%As takeaway, the following multiple research hypotheses are presented. 

This paper' analysis suggests that the core research challenge lies not in marginally improving generation accuracy, but in integrating contextual awareness, iterative refinement, methodological rigor, and reproducibility into LLM-based BPM systems.

As takeaway, the following research hypotheses are presented. 

H1: fine-tuning a SLM with complex and well-formed business processes, in a machine interpretation format, is better than the current approaches.

H2: fine-tuning a LLM with complex and well-formed business processes, in a machine interpretation format, is better than the current approaches. 

H3: SLM with no fine-tuning but co-adjuvated with RAG containing complex and well-formed business processes, in a machine interpretation format, is better than the current approaches and previous hypothesis.

H4: SLM with fine-tuning but co-adjuvated with RAG containing complex and well-formed business processes, in a machine interpretation format, is better than the current approaches and previous hypothesis.

H5: LLM with no fine-tuning but co-adjuvated with RAG containing complex and well-formed business processes, in a machine interpretation format, is better than the current approaches and previous hypothesis.

H6: LLM with fine-tuning but co-adjuvated with RAG containing complex and well-formed business processes, in a machine interpretation format, is better than the current approaches and previous hypothesis.

%%%%%%%%%%%%%%%%%
% Conclusion %
%%%%%%%%%%%%%%%%%
\section{Conclusion}
\label{sec.conclusions}

This article presented a literature review of AI-driven approaches for transforming natural language into BPMN and related process models, with a particular focus on the emerging role of LLMs. By analyzing the literature, the study mapped the main methodological approaches, examined how LLMs are currently integrated into text-to-process-model pipelines, and reviewed the evaluation strategies adopted in recent research.

The analysis shows that LLMs have significantly expanded the capabilities of automated process model generation. Compared with earlier NLP- and rule-based approaches, LLM-based methods enable more flexible interpretation of natural language descriptions, support the integration of heterogeneous information sources, and facilitate interactive modeling scenarios. However, the literature also reveals persistent limitations. Challenges related to semantic correctness, hallucinations, and structural inconsistencies remain common, particularly when dealing with complex control-flow structures or ambiguous descriptions. Moreover, evaluation practices remain fragmented, with results often depending on constrained benchmarks, subjective expert assessments, or case-specific experimental setups. It is important to note that this research area is still relatively recent, that few bibliographic references exist, and many of the existing approaches remain exploratory, which partly explains the diversity of methodologies and evaluation strategies observed in the literature.

Beyond these technical challenges, the review highlights broader methodological issues. Reproducibility is difficult by the evolving nature of LLM-based systems, and the limited availability of realistic datasets constrains the external validity of many experimental results. At the same time, many approaches treat process modeling primarily as a transformation problem, focusing on one-shot model generation rather than supporting the inherently iterative and interpretative nature of real-world process modeling activities.

Overall, these observations suggest that advancing the field requires moving beyond isolated improvements in model generation accuracy. Future research should focus on architectures that incorporate contextual knowledge, support iterative human–AI collaboration, and rely on more robust and transparent evaluation protocols. In this regard, approaches based on RAG combined with mechanisms for preserving contextual continuity across modeling iterations, and comprehensive evaluation frameworks represent promising directions for future investigation.

By synthesizing the current state of the art and critically analyzing its limitations, this article aims to provide a structured overview of existing research while supporting a broader discussion that opens new perspectives for the development of LLM-assisted business process modeling.

%%%%%%%%%%%%%%%%
% BIBLIOGRAPHY %
%%%%%%%%%%%%%%%%

\printbibliography

@techreport{Kitchenham2007,
  author      = {Kitchenham, Barbara and Charters, Stuart},
  title       = {Guidelines for Performing Systematic Literature Reviews in Software Engineering},
  institution = {EBSE Technical Report},
  number      = {EBSE-2007-01},
  year        = {2007},
  address     = {Keele University, UK}
}

@inproceedings{SchulerAlpers2024,
  author    = {Sch{\"u}ler, Selina and Alpers, Sascha},
  title     = {State of the Art: Automatic Generation of Business Process Models},
  booktitle = {Business Process Management Workshops},
  editor    = {De Weerdt, Jochen and Pufahl, Luise},
  publisher = {Springer Nature Switzerland},
  address   = {Cham, Switzerland},
  pages     = {161--173},
  year      = {2024},
  doi       = {10.1007/978-3-031-50974-2_13}
}

@inproceedings{Hafner2025,
  author    = {Hafner, Alina and Wittges, Holger and Rinderle-Ma, Stefanie},
  title     = {GenAI in Business Process Management: A Systematic Review of the Current State},
  booktitle = {Proceedings of the 31st Americas Conference on Information Systems (AMCIS)},
  year      = {2025},
  address   = {Montréal, Canada},
  publisher = {Association for Information Systems},
  url       = {https://aisel.aisnet.org/amcis2025/sig_svc/sig_svc/9}
}

@inproceedings{Qiu2025,
  author    = {Qiu, Han and Chen, Hongyun and Zou, Baoyu and Dong, Zizheng},
  title     = {Intelligent Design and Implementation of Government Affairs Processes Driven by Large Language Models},
  booktitle = {Proceedings of the 2024 International Conference on Artificial Intelligence of Things and Computing (AITC 2024)},
  year      = {2025},
  pages     = {70--79},
  doi       = {10.1145/3708282.3708296}
}

@inproceedings{Ajmal2024,
  author    = {Ajmal, Farhath and Wijekoon, Poorna and Dhanamina, Haritha and Ravishan, Yasiru and Nawinna, Dasuni and Attanayaka, Buddhima},
  title     = {Automated BPMN Diagram Generation},
  booktitle = {Proceedings of the 6th International Conference on Advancements in Computing (ICAC 2024)},
  year      = {2024},
  pages     = {7--12},
  doi       = {10.1109/ICAC64487.2024.10851120}
}

@inproceedings{Kogler2024,
  author    = {Kogler, Philipp and Chen, Wei and Falkner, Andreas and Haselb{\"o}ck, Alois and Wallner, Stefan},
  title     = {Modelling Engineering Processes in Natural Language: A Case Study},
  booktitle = {ACM International Conference Proceeding Series},
  year      = {2024},
  pages     = {170--178},
  doi       = {10.1145/3646548.3672584}
}

@inproceedings{Ivanchikj2020,
  author    = {Ivanchikj, Ana and Serbout, Souhaila and Pautasso, Cesare},
  title     = {From Text to Visual BPMN Process Models: Design and Evaluation},
  booktitle = {Proceedings of the 23rd ACM/IEEE International Conference on Model Driven Engineering Languages and Systems (MODELS 2020)},
  year      = {2020},
  pages     = {229--239},
  doi       = {10.1145/3365438.3410990}
}

@inproceedings{Ziche2024,
  author    = {Ziche, Clara and Apruzzese, Giovanni},
  title     = {LLM4PM: A Case Study on Using Large Language Models for Process Modeling in Enterprise Organizations},
  booktitle = {Lecture Notes in Business Information Processing},
  volume    = {527},
  year      = {2024},
  pages     = {472--483},
  doi       = {10.1007/978-3-031-70445-1_35}
}

@article{Li2025,
  author  = {Li, Shan and Liu, Zesan and Liu, Xinyi and Chen, Gang and Zhang, Yongjian and Ling, Haojie and Huang, Xing},
  title   = {Structured Extraction and BPMN Generation From Chinese Procedural Texts via Hybrid Attention-Based Domain Language Models},
  journal = {IEEE Access},
  volume  = {13},
  year    = {2025},
  pages   = {166241--166253},
  doi     = {10.1109/ACCESS.2025.3611811}
}

@article{Kourani2025,
  author  = {Kourani, Humam and Berti, Alessandro and Schuster, Daniel and van der Aalst, Wil M. P.},
  title   = {Evaluating Large Language Models on Business Process Modeling: Framework, Benchmark, and Self-Improvement Analysis},
  journal = {Software and Systems Modeling},
  year    = {2025},
  doi     = {10.1007/s10270-025-01318-w}
}

@inproceedings{Neuberger2025,
  author    = {Neuberger, Julian and Ackermann, Lars and van der Aa, Han and Jablonski, Stefan},
  title     = {A Universal Prompting Strategy for Extracting Process Model Information from Natural Language Text Using Large Language Models},
  booktitle = {Lecture Notes in Computer Science},
  volume    = {15238},
  year      = {2025},
  pages     = {38--55},
  doi       = {10.1007/978-3-031-75872-0_3}
}

@inproceedings{Bellan2022,
  author    = {Bellan, Patrizio and Dragoni, Mauro and Ghidini, Chiara},
  title     = {Extracting Business Process Entities and Relations from Text Using Pre-trained Language Models and In-Context Learning},
  booktitle = {Lecture Notes in Computer Science},
  volume    = {13585},
  year      = {2022},
  pages     = {182--199},
  doi       = {10.1007/978-3-031-17604-3_11}
}

@article{Tsaneva2024,
  author  = {Tsaneva, Stefani and Sabou, Marta},
  title   = {Enhancing Human-in-The-Loop Ontology Curation Results through Task Design},
  journal = {Journal of Data and Information Quality},
  volume  = {16},
  number  = {1},
  pages   = {4:1--4:25},
  year    = {2024},
  doi     = {10.1145/3626960}
}

@inproceedings{Bennoit2024,
  author    = {Bennoit, Christian and Greff, Tobias and Bajwa, Imran Abdullah and Baum, Deborah},
  title     = {Identifying Use Cases for Large Language Models in the Business Process Management Lifecycle},
  booktitle = {2024 26th International Conference on Business Informatics (CBI)},
  year      = {2024},
  pages     = {256--263},
  doi       = {10.1109/CBI62504.2024.00037},
  publisher = {IEEE}
}

@inproceedings{Buss2025,
  author    = {Buss, Alina and Kratsch, Wolfgang and Schmid, Sebastian Johannes and Wang, Hongyang},
  title     = {ProcessLLM: A Large Language Model Specialized in the Interpretation, Analysis, and Optimization of Business Processes},
  booktitle = {Lecture Notes in Business Information Processing (BPM 2024 Workshops)},
  volume    = {534},
  pages     = {221--232},
  year      = {2025},
  doi       = {10.1007/978-3-031-78666-2_17}
}

@article{Sun2025,
  author    = {Sun, Xiaoxiao and Zhao, Chenying and Yu, Dongjin and Xu, Yi and Xiao, Nana},
  title     = {TeProM: A Rule-Free Method for Extracting Process from Complex Text with Enhanced Coreference Handling},
  journal   = {Information Sciences},
  volume    = {719},
  pages     = {122451},
  year      = {2025},
  doi       = {10.1016/j.ins.2025.122451}
}

@inproceedings{Nivon2025,
  author    = {Nivon, Quentin and Salaün, Gwen and Lang, Frédéric},
  title     = {GIVUP: Automated Generation and Verification of Textual Process Descriptions},
  booktitle = {33rd ACM International Conference on the Foundations of Software Engineering (FSE Companion '25)},
  year      = {2025},
  pages     = {1119--1123},
  doi       = {10.1145/3696630.3728593},
  publisher = {ACM}
}

@techreport{Bettencourt2026,
  author      = {XX, XX},
  title       = {{Final Data Extraction and Manipulation Sheet in my LR}},
  year        = {2026},
  month       = jan,
  language    = {en},
  institution = {YY},
  copyright   = {metadata-only-access},
  url         = {https://scholar.tecnico.ulisboa.pt/api/records/xzohbq2uqkyyp4Ha8jPGTCzsPa2AWkUAUR-y/file/5e07ad7654d2bee72dd098a684574504c3c2348fba12ec138dd576df241c8df6.xlsx}
}

@mastersthesis{Azeredo2025,
  author       = {Azeredo, Francisco},
  title        = {Enhancing Semantic Search with NLP for Secure and Scalable Document Retrieval: A Metadata-Enriched, ACL-Aware Approach},
  school       = {Instituto Superior T{\'e}cnico, University of Lisbon},
  year         = {2025},
  type         = {Master's Thesis},
}

@inproceedings{Kourani2024,
  author    = {Kourani, Humam and Berti, Alessandro and Schuster, Daniel and van der Aalst, Wil M. P.},
  title     = {Process Modeling with Large Language Models},
  booktitle = {Lecture Notes in Business Information Processing},
  volume    = {511},
  year      = {2024},
  pages     = {229--244},
  doi       = {10.1007/978-3-031-61007-3_18}
}

@book{Dumas2018,
  author    = {Dumas, Marlon and La Rosa, Marcello and Mendling, Jan and Reijers, Hajo A.},
  title     = {Fundamentals of Business Process Management},
  edition   = {2},
  publisher = {Springer},
  address   = {Berlin, Heidelberg},
  year      = {2018},
  doi       = {10.1007/978-3-662-56509-4}
}

@article{Hoerner2026,
  author  = {H{\"o}rner, Luca Franziska and M{\"o}ller, Maximilian and Reichert, Manfred},
  title   = {Automatically Generating BPMN 2.0 Process Models from Natural Language Process Descriptions: Challenges, Framework, Quality Assessment},
  journal = {Software and Systems Modeling},
  year    = {2026},
  note    = {Accepted after two revisions for a Special Issue},
}

@misc{Lauer2026,
  author  = {Lauer, Chantale and Pfeiffer, Peter and Rombach, Alexander and Mehdiyev, Nijat},
  title   = {Assessing the Business Process Modeling Competences of Large Language Models},
  year    = {2026},
  note    = {arXiv preprint},
  url     = {https://arxiv.org/abs/2601.21787}
}

@misc{winograd1987understanding,
  title={On understanding computers and cognition: A new foundation for design: A response to the reviews},
  author={Winograd, Terry and Flores, Fernando},
  year={1987},
  publisher={Elsevier}
}

@book{gianola2023verification,
  title={Verification of Data-Aware Processes via Satisfiability Modulo Theories},
  author={Gianola, Alessandro},
  year={2023},
  publisher={Springer}
}

@inproceedings{calvanese2019formal,
  title={Formal modeling and SMT-based parameterized verification of data-aware BPMN},
  author={Calvanese, Diego and Ghilardi, Silvio and Gianola, Alessandro and Montali, Marco and Rivkin, Andrey},
  booktitle={International Conference on Business Process Management},
  pages={157--175},
  year={2019},
  organization={Springer}
}

@misc{bpmn2014,
    title = {Business Process Model and Notation - BPMN - 2.0.2},
    author = {OMG},
    organization = {Object Management Group},
    year = {2014},
    url = {https://www.omg.org/spec/BPMN},
    note = {specification (ISO/IEC 19510:2013)}
}

@article{dijkman2008semantics,
  title={Semantics and analysis of business process models in BPMN},
  author={Dijkman, Remco M and Dumas, Marlon and Ouyang, Chun},
  journal={Information and Software technology},
  volume={50},
  number={12},
  pages={1281--1294},
  year={2008},
  publisher={Elsevier}
}

@inproceedings{fickinger2013construct,
  title={Construct redundancy in process modelling grammars: Improving the explanatory power of ontological analysis},
  author={Fickinger, Tobias and Recker, Jan},
  booktitle={Proceedings of the 21st European Conference on Information Systems},
  pages={1--12},
  year={2013},
  organization={Association for Information Systems (AIS)}
}

@article{laue2010visualization,
  title={Visualization of business process modeling anti patterns},
  author={Laue, Ralf and Awad, Ahmed},
  journal={Electronic Communications of the EASST},
  volume={25},
  year={2010}
}

@article{dumas2023ai,
  title={AI-augmented business process management systems: a research manifesto},
  author={Dumas, Marlon and Fournier, Fabiana and Limonad, Lior and Marrella, Andrea and Montali, Marco and Rehse, Jana-Rebecca and Accorsi, Rafael and Calvanese, Diego and De Giacomo, Giuseppe and Fahland, Dirk and others},
  journal={ACM Transactions on Management Information Systems},
  volume={14},
  number={1},
  pages={1--19},
  year={2023},
  publisher={ACM New York, NY}
}

@inproceedings{genon2010analysing,
  title={Analysing the cognitive effectiveness of the BPMN 2.0 visual notation},
  author={Genon, Nicolas and Heymans, Patrick and Amyot, Daniel},
  booktitle={International conference on software language engineering},
  pages={377--396},
  year={2010},
  organization={Springer}
}

@book{dietz2024enterprise,
  title={Enterprise ontology: A human-centric approach to understanding the essence of organisation},
  author={Dietz, Jan LG and Mulder, Hans BF},
  year={2024},
  publisher={Springer Nature}
}

@article{beerepoot2023biggest,
  title={The biggest business process management problems to solve before we die},
  author={Beerepoot, Iris and Di Ciccio, Claudio and Reijers, Hajo A and Rinderle-Ma, Stefanie and Bandara, Wasana and Burattin, Andrea and Calvanese, Diego and Chen, Tianwa and Cohen, Izack and Depaire, Beno{\^\i}t and others},
  journal={Computers in Industry},
  volume={146},
  pages={103837},
  year={2023},
  publisher={Elsevier}
}

@article{guerreiro2021conceptualizing,
  title={Conceptualizing on dynamically stable business processes operation: a literature review on existing concepts},
  author={Guerreiro, S{\'e}rgio},
  journal={Business Process Management Journal},
  volume={27},
  number={1},
  pages={24--54},
  year={2021},
  publisher={Emerald Publishing Limited}
}

@article{lopes2023assessing,
  title={Assessing business process models: a literature review on techniques for BPMN testing and formal verification},
  author={Lopes, Tom{\'a}s and Guerreiro, S{\'e}rgio},
  journal={Business Process Management Journal},
  volume={29},
  number={8},
  pages={133--162},
  year={2023},
  publisher={Emerald Publishing Limited}
}

@article{milani2025business,
  title={Business process improvement opportunities and redesign options: a systematic literature review},
  author={Milani, Fredrik and Lashkevich, Katsiaryna},
  journal={Business Process Management Journal},
  pages={1--21},
  year={2025},
  publisher={Emerald Publishing Limited}
}

%%%%%%%%%%%%
% APPENDIX %
%%%%%%%%%%%%
% \clearpage % <-- ACTION: Comment the following blocks if you dont need appendixes
% \appendix
% \input{Sections/AppendixA.tex}

%%%%%%%%%%%%%
%    END    %
%%%%%%%%%%%%%

\end{document}